\documentclass[journal]{IEEEtran}
\pdfoutput=1
\usepackage{hyperref}
\usepackage{scalerel}
\usepackage{tikz}
\usetikzlibrary{svg.path}
\usepackage{xcolor}
\usepackage{cite}
\usepackage{algorithm}
\usepackage{algpseudocode}
\usepackage{stackengine}
\usepackage[LGRgreek]{mathastext}
\usepackage{amsmath,amssymb,amsfonts}
\usepackage{graphicx}
\usepackage{textcomp}
\usepackage{xcolor}
\usepackage[utf8]{inputenc}
\usepackage{multirow}
\usepackage{array}
\usepackage{booktabs}
\usepackage{soul}
\usepackage{hyperref}
\usepackage{lineno}

\definecolor{orcidlogocol}{HTML}{A6CE39}

\tikzset{
  orcidlogo/.pic={
    \fill[orcidlogocol] svg{M256,128c0,70.7-57.3,128-128,128C57.3,256,0,198.7,0,128C0,57.3,57.3,0,128,0C198.7,0,256,57.3,256,128z};
    \fill[white] svg{M86.3,186.2H70.9V79.1h15.4v48.4V186.2z}
                 svg{M108.9,79.1h41.6c39.6,0,57,28.3,57,53.6c0,27.5-21.5,53.6-56.8,53.6h-41.8V79.1z M124.3,172.4h24.5c34.9,0,42.9-26.5,42.9-39.7c0-21.5-13.7-39.7-43.7-39.7h-23.7V172.4z}
                 svg{M88.7,56.8c0,5.5-4.5,10.1-10.1,10.1c-5.6,0-10.1-4.6-10.1-10.1c0-5.6,4.5-10.1,10.1-10.1C84.2,46.7,88.7,51.3,88.7,56.8z};
  }
}

\newcommand\orcidicon[1]{\href{https://orcid.org/#1}{\mbox{\scalerel*{
\begin{tikzpicture}[yscale=-1,transform shape]
\pic{orcidlogo};
\end{tikzpicture}
}{|}}}}

\begin{document}
\title{
Weave and Conquer: A Measurement-based Analysis of Dense Antenna Deployments}
\author{Andrea P. Guevara \orcidicon{0000-0002-7496-4772}, ~\IEEEmembership{Student Member,~IEEE,}
        Sibren De Bast \orcidicon{0000-0002-9456-4709}, ~\IEEEmembership{Student Member,~IEEE,}
        Sofie Pollin \orcidicon{0000-0002-1470-2076}, ~\IEEEmembership{Senior Member,~IEEE}}
\maketitle

\begin{abstract}
Massive MIMO is bringing significant performance improvements in the context of outdoor macrocells, such as favourable propagation conditions, spatially confined communication, high antenna gains to overcome pathloss, and good angular localisation. In this paper we explore how these benefits scale to indoor scattering-rich deployments based on a dense indoor measured Massive MIMO dataset. First, we design and implement three different and relevant topologies to position our 64 antennas in the environment: \emph{Massive MIMO, RadioStripes} and \emph{RadioWeaves} topologies. Second, we measure 252004 indoor channels for a $3x3m^2$ area for each topology, using an automated user-positioning and measurement system. Using this dense dataset, we provide a unique analysis of system level properties such as pathloss, favourable propagation, spatial focusing and localisation performance. Our measurement-based analyses verify and quantify that distributing the antennas throughout the environment results in an improved propagation fairness, better favourable propagation conditions, higher spatial confinement and finally a high localisation performance.
The dataset is publicly available and can serve as a reference database for benchmarking of future indoor communication systems and communication models. We outline the implementation challenges we observed, and also list diverse R\&D challenges that can benefit from using this dataset as a benchmark. 

\end{abstract}


\IEEEpeerreviewmaketitle

\section{Introduction}
As network densification continues, Massive MIMO systems will surround us outdoor as well as indoor. Subsequently, the key question is how these multiple antennas should be woven in our environment. The concept beyond massive MIMO in indoor scenarios has been studied before under multiple terms as \textit{Large Intelligent Surfaces} (LIS) \cite{Hu2017} or \textit{e-walls} \cite{SalarRahimi}. In a LIS system, thousands of identical antennas are deployed in a grid-fashion over surfaces like walls, those antennas have the same distance between each other and are arranged in a single or multiple arrays. As the number of antennas grows, a super-directivity is created towards the intended users \cite{Williams2019}, therefore, increasing the system capacity \cite{Hu2018}. On the other hand, the vision of the Ericsson team is related to the distribution of antennas next to each other to create a so called \textit{RadioStripe} \cite{Ericsson}. The antennas are attached to dedicated cables, which can easily be deployed on top of any surface, both indoor and outdoor. This creates a flexible and highly distributed massive MIMO system. 
\begin{figure}[!ht]
    \centering
    \includegraphics[width=\linewidth]{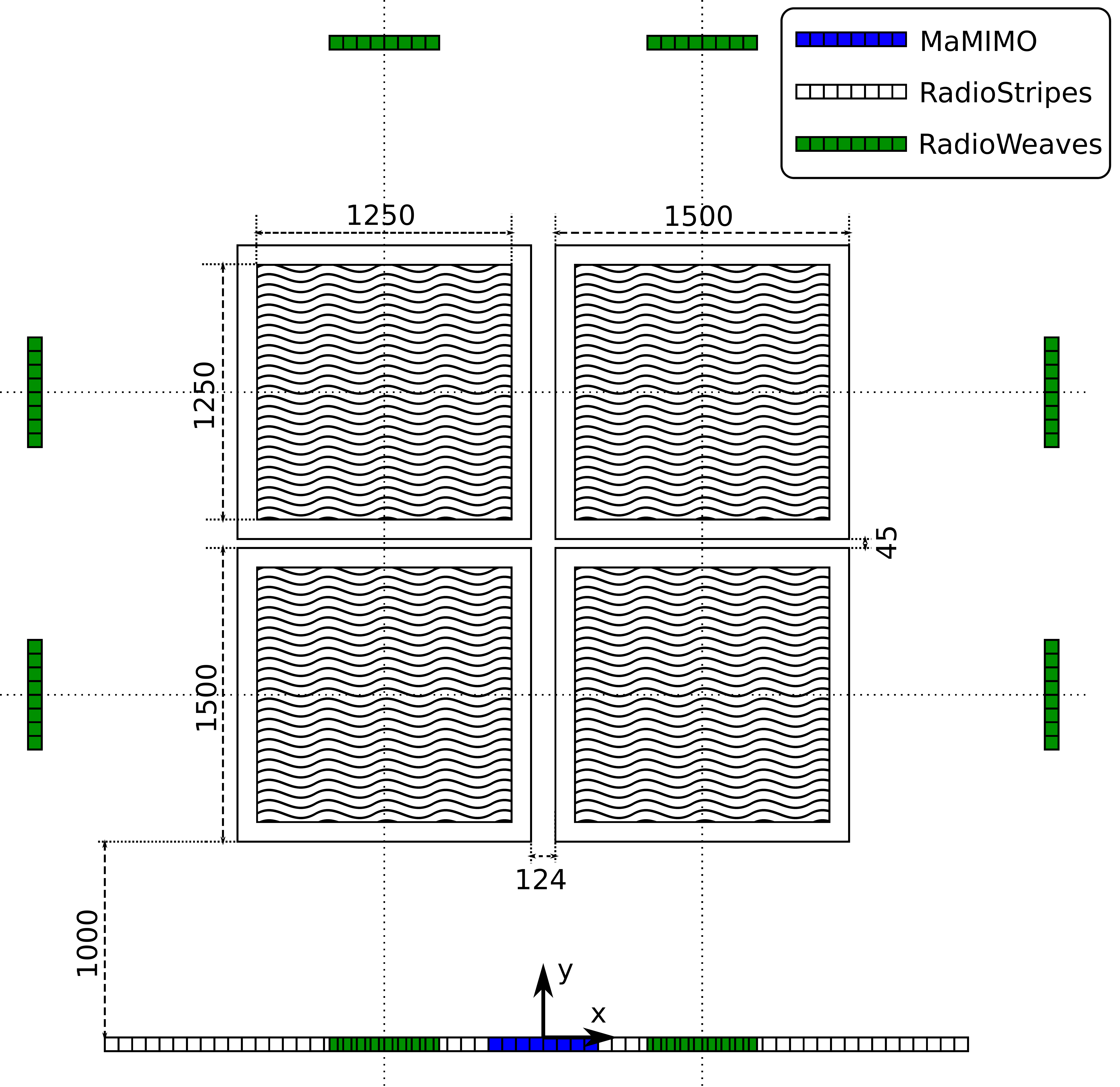}
    \caption{The three different methods for \emph{weaving} the 64 antennas in the environment in our experiments are illustrated using green, blue or white boxes. A traditional \emph{Massive MIMO} deployment centralises all elements in a compact uniform rectangular array, shown in blue, while the \emph{RadioStripes} topology results in a long uniform linear array deployment, depicted in white.  
    The antennas can also be distributed in scattered clusters over the environment, this is illustrated in green as \textit{RadioWeaves}. The dimensions of the measured space and exact locations of the 64 antenna elements are given in mm. The waved region denotes the area where the users are located during the measurements. 
    }
    \label{fig:scenario}
\end{figure}

The first conceptual paper introducing the \emph{RadioWeaves} concept was \cite{VanderPerre2020}. The concept is an evolution of other pioneering contributions in the domain of Massive MIMO and Cell-free architectures,
 based on a distributed radio and computing infrastructure. In \emph{RadioWeaves}, antennas are essentially distributed, enabling an increase of the number of antennas. This creates both statistically favourable propagation conditions as well as favourable coverage conditions, which means that every position is in the main lobe and in close proximity of at least one antenna element. Furthermore, by adding distributed compute power, it is envisioned that new intelligent applications can be supported by such future networks operating in, for example, crowd scenarios or factory environments. 

In this paper, we focus on  
 the antenna distribution in indoor scenarios. Based on the unique large and flexible KU Leuven Massive MIMO testbed \cite{bast2019csibased,Guevara2020}, we create a large and dense indoor channel database that can be used to evaluate favourable coverage and favourable propagation conditions for multiple antenna deployment strategies. Fig. \ref{fig:scenario} shows the different indoor antenna distributions that will be studied in this paper: Centralised \textit{Massive MIMO}, \textit{RadioStripes} and \textit{RadioWeaves}.

For future indoor applications, other metrics than coverage or spectral efficiency will become more and more important. While 5G already introduces reliability and latency, we envision also other primitives such as localisation accuracy or even power focusing 
in the spatial domain. 
However, to have a wide understanding of the wireless channel and 
suitable signal processing techniques, different experiments in a variety of scenarios must be carried out. 

Channel collection experiments in real scenarios are limited due to the 
manual effort that such simple experiments entail
, and the lack of funding to deploy massive MIMO testbeds. Moreover, the datasets that do get created are seldom shared with the public. Furthermore, there is a significant imbalance in the number of theoretical versus experimental publications, and it is not possible for the small experimental community to keep up with the theoretical community to sufficiently validate the assumptions made in their work. 
Nevertheless, some experimental Massive MIMO papers exist in the literature, these study among others wireless channel characteristics as: channel hardening \cite{Gunnarsson18}, temporal correlation \cite{harris2017temporal}, and antenna power contribution for two centralised antenna topologies \cite{Gao2015}. However, none compare multiple antenna deployment strategies while relying on a super dense channel measurement dataset. 

\subsection{Contributions}
This work provides an experimental evaluation of three indoor antenna distribution strategies as an aspect of the indoor \emph{RadioWeaves} concept. The concrete contributions are:

\begin{enumerate}
\item A comparisons of three relevant strategies to weave 64 antennas in a single room: Centralised \emph{Massive MIMO} as a uniform rectangular array, \emph{RadioStripes} as a uniform linear array and finally \emph{RadioWeaves} as a distributed deployment of 8 sub-arrays. 
\item A dense data collection strategy and dataset for these three deployment strategies, consisting of 252004 points for each scenario. Both a LoS and a NLoS scenario are considered for \textit{Massive MIMO}.
 The dataset is publicly available \href{https://homes.esat.kuleuven.be/~sdebast/measurements/measurements_boardroom.html}{here}.
\item A comparison of typical connectivity KPI such as favourable coverage (Received Signal Strength) and propagation conditions (Position Correlation Function). 
\item A study of the power focusing performance and interference statistics when using the simple MRT for a range of user deployment scenarios. 
\item Insight in the use of future dense networks following the \emph{RadioWeaves} concept for other services such as localisation, following our earlier work in \cite{bast2019csibased}.
\end{enumerate}
As the dataset is publicly available, it can be used by the community for further studies beyond those described in this paper.

\section{Indoor Modular Massive MIMO Measurements}
In order to couple the theoretic understanding of Massive MIMO systems to the practical reality, an extensive measurement campaign was carried out as the foundation of this study. These measurements used the flexible KU Leuven MaMIMO testbed located at the department of electrical engineering (ESAT). This testbed is equipped with 64 antennas at the base station (BS) and can serve simultaneously up to 12 users. The KU Leuven testbed is TDD-LTE based and controlled via the MIMO Application Framework of National Instruments; the main parameters are detailed in \cite{lutherNI}. For the measurements, we use 64 patch antennas at the base station \cite{Chen2017} and a dipole antenna at the user, using a centre frequency of 2.61 GHz.

\subsection{Antenna Deployments}
One of the main features of the KU Leuven massive MIMO system are the 64 modular antenna elements, which can be arranged in multiple complex array configurations. To emulate different future indoor massive MIMO technologies, three of these configurations were deployed for this study.
\begin{itemize}
        \item Standard centralised \textbf{Massive MIMO} as a Uniform Rectangular Array (URA): All the antenna elements are deployed in a centralised manner, to form an $8\times8$ antenna array with a total size of $560\times560$ mm. This array was placed at the centre in front of the users (see blue boxes in Fig.  \ref{fig:scenario}). For this particular topology only, a non-Line-of-Sight \textbf{NLoS} scenario was considered, placing a metal plane in front of the antenna array at a distance of 500 mm.
        \item \textbf{RadioStripes} as a Uniform Linear Array (ULA): The  \emph{RadioStripes} concept proposes a large number of antennas deployed in a linear array. To measure this topology, the 64-antenna elements were deployed in a single line with a length of 4480 mm. This is depicted in Fig. \ref{fig:scenario} using white boxes. This array is located at a distance of 1 m from the positioners, and both centres are aligned in the x-direction.
        \item \textbf{RadioWeaves} as Distributed ULA (D-ULA): In this case, 8 sub-arrays of 8 antennas are distributed in the room, uniformly around the measured area, shown by the green boxes in Fig. \ref{fig:scenario}.
\end{itemize}

All the antennas were placed at a height of 1 m, in the case of the centralised Massive MIMO scenario, the bottom antennas of the rectangular array were positioned at a height of 1 m.
\subsection{UE Deployment and Channel Measurements}
During the measurement campaign four users were automatically and synchronously moved over a grid in an area of $1250\times1250$ mm, illustrated by the wavey area in Fig. \ref{fig:scenario}. 
Every 5 mm, the movement or the four users was paused for a static channel collection for the four users. 
Due to the short sub-cm measurement granularity, we call this an ultra-dense dataset. 
For each location $\textit{l} \in [ 1 \dots \textit{L}] $ with \textit{L =} 252004, the channel state information (CSI) can be noted as a matrix $\mathbf{H}_\mathit{l} \in \mathbb{C}^{64 \times 100}$, with the first dimension spanning the 64 BS antennas and the 100 measured 
resource blocks spanning the second dimension.

\subsection{Dataset Limitations}
It is worth to mention that all the cables between each patch antenna and their RF equipment, are exactly equal in 
length and characteristics, although each cable is subject to small phase and amplitude variations. 
In our dataset and measurements, we did not do a calibration across antennas, as the phase and gain errors are small and moreover constant across each measured position and antenna topology. 
As a result, the absolute values of channel gain and phases are not obtained with the dataset, and only relative comparisons of topologies, precoders or locations can be done with the current dataset. In addition to the uncalibrated phase mismatch of the cables, the BS-UE synchronisation is done over the air, using the traditional LTE synchronisation method, and is not precise enough to obtain the channel phase.

\section{Favourable coverage conditions} \label{RSSI}
Distributing the antennas over the environment is expected to result in favourable propagation conditions for the full coverage zone. This is due to the diverse placement and orientation of the antenna arrays and elements. 
In \cite{Gao2015} it was demonstrated experimentally that not all antenna elements contribute equally to the received signal strength. This was attributed to both channel and antenna pattern variations. 
It was shown that, for a cylindrical and linear array, a proper antenna selection improves the system performance. Based on this premise, and knowing that the antenna configuration influences the channel correlation, the normalised uplink Received Signal Strength (RSS) is collected and analysed for all the different user locations and antenna configurations. 
The RSS is essentially the mean uplink received power averaged over all 64 antennas and 100 resource blocks. 

\begin{figure}
    \centering
    \includegraphics[width=\linewidth]{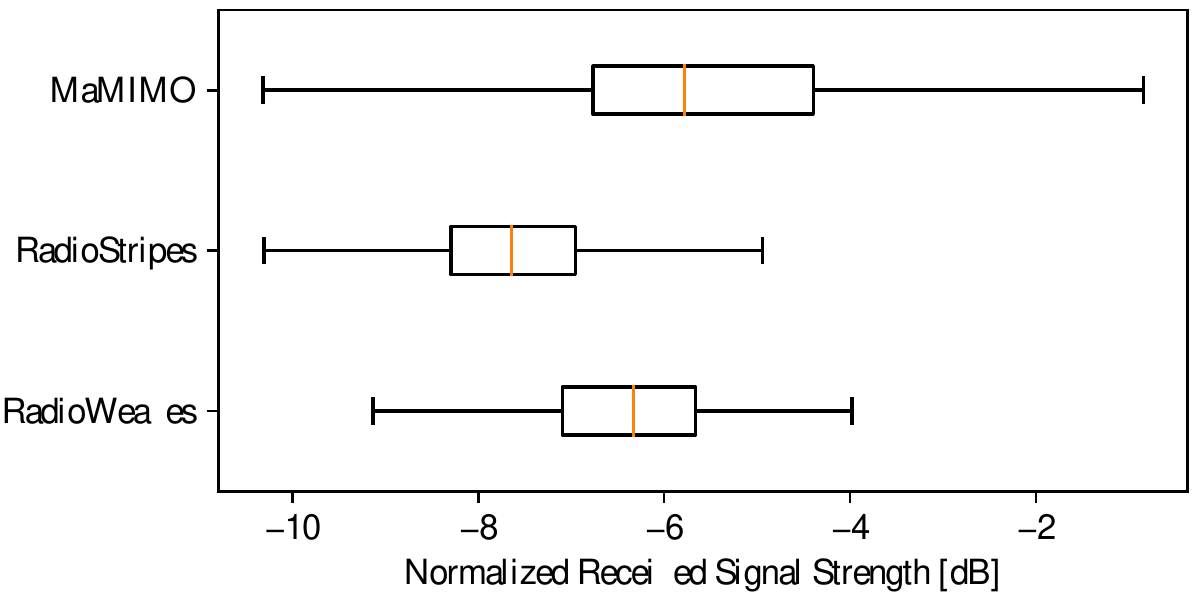}
    \caption{Weaving the radios in the environment can maximise the probability of being close to one or more arrays and be in a good view of at least one antenna element. This figures quantifies the normalised mean RSSI obtained with 64 antenna elements and MRT. While \emph{Massive MIMO} can result in a very large signal for some (nearby) users, the worst user also has a 10dB lower signal already in such a small area. The \emph{RadioStripes}, by design, cannot achieve maximal array gain in the near field as not all antennas contributed equally. Finally, the \emph{RadioWeaves} distributed architecture, which means the antennas are maximally woven in the environment, results clearly in the most fair system design, although also in this scenario not all antennas contribute equally and the highest possible array gain is never achieved.} 
    \label{fig:RSS}
\end{figure}

Favourable coverage would be obtained when all possible \textit{L} user locations have a high mean RSS and high minimal RSS, this would ensure a good coverage for even the worst locations. 
As expected, when all antennas are centralised in a single place (as the \textit{Massive MIMO} configuration), the mean coverage or RSS will be high for the users close to the array. This type of antenna configuration creates an uneven 
coverage distribution, which can easily be seen as a large spread of the values in Fig. \ref{fig:RSS}. In our experiment the distance between the \textit{centralised Massive MIMO} and the closest user is $1 m$ and around $4 m$ for the furthest user, which represents a variation of 9dB of the mean RSS.

The \textit{RadioStripes} distribution is the preferred topology by most of the theoretical analyses, due to the simplicity to model the channel in the far field. However, when considering an indoor \textit{RadioStripes} deployment, the RSS distribution of this topology
is more complex to understand. As the linear array is spread maximally in the room,  
the energy is distributed fairer than it was in the case of the \textit{centralised Massive MIMO} configuration. 
Fig. \ref{fig:RSS} illustrates that the \emph{RadioStripes} indeed have a lower spread in RSS, however, the mean RSS over the entire coverage region is lower. In fact, the \emph{RadioStripes} deployment results overall in a lower received signal, which is caused by the fact that the users are all located in the near field of the antenna array, i.e. no location is in the main lobe of all 64 antenna elements. 

For the \textit{RadioWeaves} topology, the probability of any user to be closer to any of the antennas is higher, in consequence the mean RSS for all the positions is higher than for the \textit{RadioStripes} topology and closer to the mean RSS of the \textit{Massive MIMO} topology. In addition, the distribution of the power is more homogeneous, resulting in more favourable coverage as the lowest measured RSS in the area has increased and the RSS spread is reduced to only 5dB. 

\section{Favourable Propagation Conditions}

Favourable propagation conditions mean that the channels experienced by users at different locations become statistically independent. As the number of antennas in large Massive MIMO arrays increases, propagation becomes more and more favourable and users can be separated more easily 

In \cite{Gao2015} the impact of two centralised antenna configurations, a linear and cylindrical array, on the channel correlation is studied experimentally for an outdoor scenario and a limited number of spatial locations. Here it is confirmed that as the number of active elements increases,  the channel between users become orthogonal, and the linear array a better favourable propagation compared to the cylindrical array of the same size.  
This lead us to the following question: \textit{Is this behaviour similar for an indoor scenario and what antenna deployment strategy is optimal?}

\subsection{Varying the weaving strategy}
\begin{figure}[!t]
    \centering
    \includegraphics[width=\linewidth]{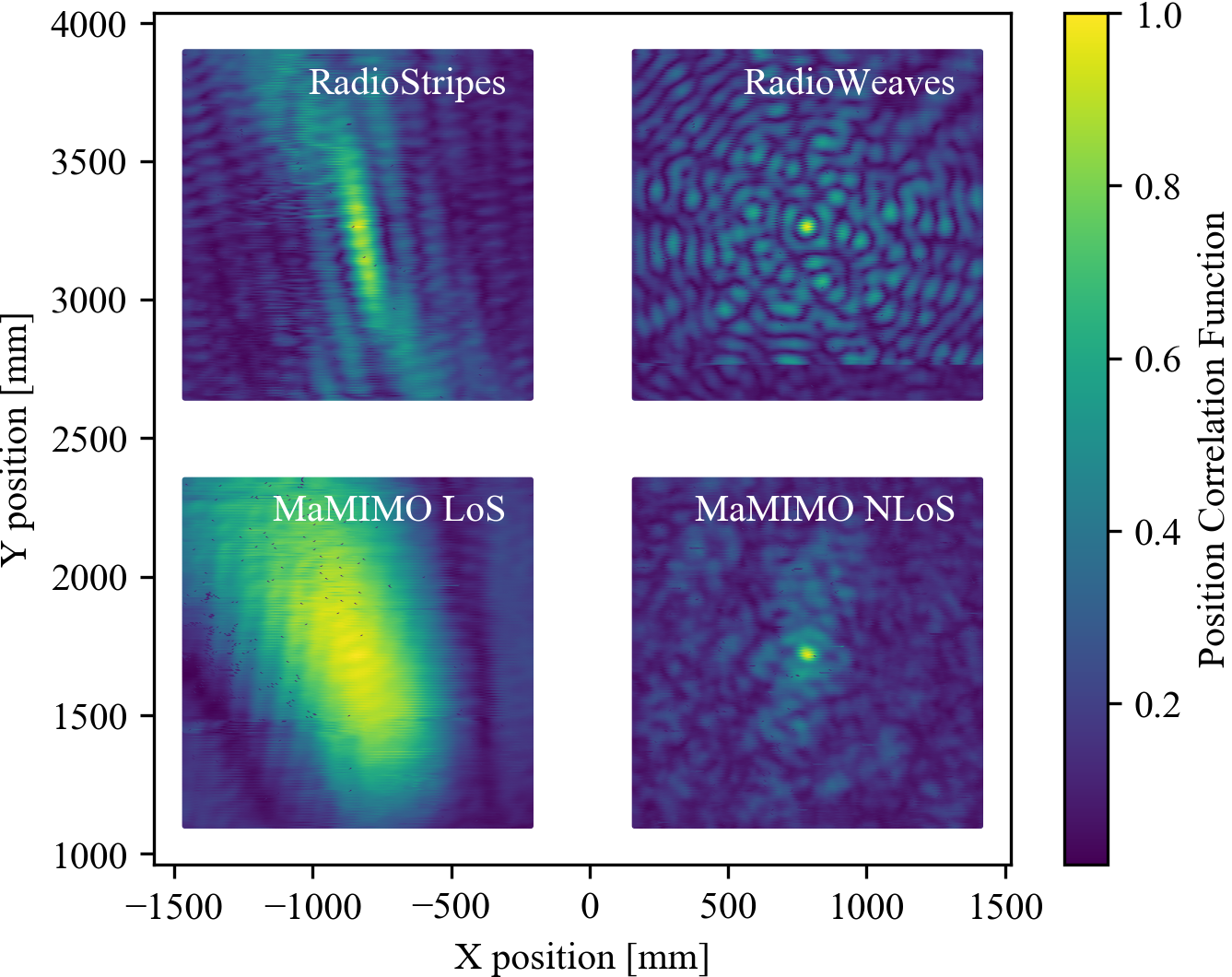}
    \caption{
    This figure illustrates the position correlation function for four different scenarios: i) \textit{MaMIMO under} LoS conditions; ii) \textit{MaMIMO} under nLoS conditions; iii) \textit{RadioStripes}; and iv) \textit{RadioWeaves}. Each positioner is used to present one of these scenarios and the PCF is calculated in relation to the centre position of each positioner.}
 
    \label{fig:scf}
\end{figure}

We study the position correlation function (PCF) empirically using our ultra-dense channel database. For each of the four positioners, the correlation of the channel of each location with the channel of the central user is determined and represented in Fig. \ref{fig:scf}. The position correlation function is essentially the 
favourable propagation metric stated in \cite{Ngo2014}.  
So, when the PCF$\to 0$ then the channel 
of that position is less correlated to the reference one. 

Fig. \ref{fig:scf} compares the experimental PCF, each positioner here represents data obtained with a different antenna deployment topology. 
In the upper left square the results from the 
\emph{RadioStripes} topology show 
a highly correlated vertical area (green to yellow) around the centre of the square where the reference position is considered. Interestingly we can see that this area fades out quickly horizontally. 

The PCF of the centralised \emph{Massive MIMO} LoS deployment is shown on the bottom-left square in Fig. \ref{fig:scf}. Compared to the \emph{RadioStripes} we see that the PCF fades slower and overall more correlation is experienced in the measured plane.

Interestingly the favourable propagation conditions for a \emph{Massive MIMO} array in a NLoS scenario is improved significantly.  
 The reason behind this phenomena is the rich multipath environment created due to the metal plane, therefore, physically the waves of each antenna travel in multiple directions until they reach the user and this effect is repeated independently for every user location, creating a lower channel correlation around the reference point.

A lower PCF is also noted for the  \emph{RadioWeaves} topology, in the top-right square in Fig. \ref{fig:scf}. This is also an artificial rich multipath scenario created by the antenna distribution.  We see a wave-like PCF originating in the centre for the square.

Based on the experiments we see that the best scenario is the one that creates a rich multipath environment and leads to a lower channel correlation, therefore in LoS scenarios a \textit{RadioWeaves} antenna configuration is highly recommended.

\subsection{Scaling the number of antennas}\label{Sec. Antennas_MRT}
The results presented above based on the {PCF} can also be extended with the use of a precoder, resulting in the focusing of the intended signal on the target location. 
In this section, we focus on the simplest precoding scheme know as Maximum Ratio Transmission (MRT). 
The well known work done by Prof. Larsson et all. in \cite{Larsson2014} already presented the narrow beam created by MR as a relative strength field when a \textit{RadioStripe} is implemented using simulations.

\begin{figure}[!t]
    \centering
    \includegraphics[width=\linewidth]{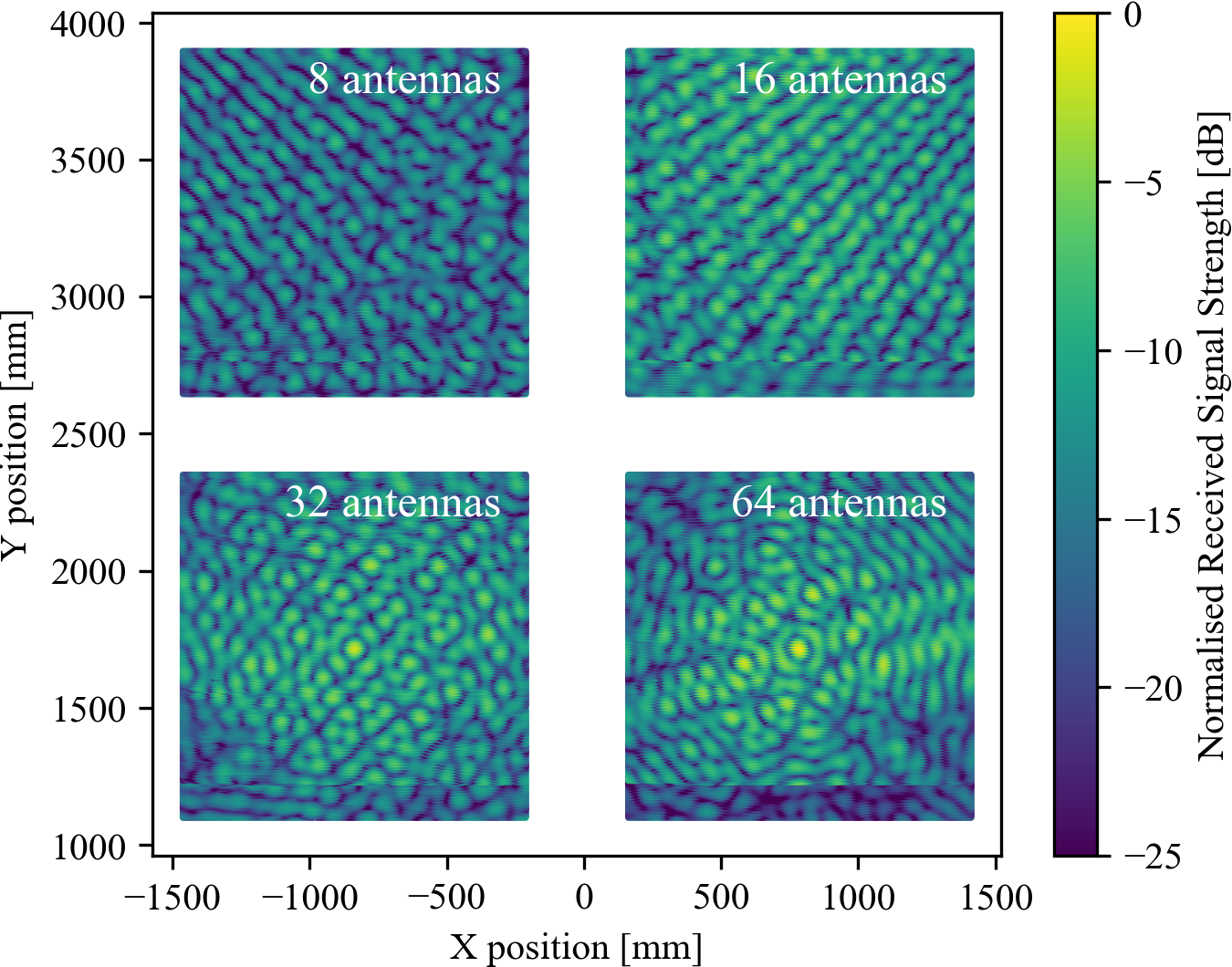}
    \caption{Normalised received signal strength as MR beamforming is applied towards the centre user of each positioner. The results in each square are presented for a different subset of antennas using the \textit{RadioWeaves} antenna configuration. As the number of antennas increases the beam surrounding the reference user becomes more directive.}
    \label{fig:Antennas}
\end{figure}

When considering \emph{RadioWeaves} it becomes more difficult to imagine
the resulting beam, and Fig.~\ref{fig:Antennas} experimentally verifies the spot focusing performance of our measured uniform distribution of the antenna sub-arrays in a room when a total of 8, 16, 32 and 64 antennas are active. It is worth mentioning that the MRT precoding vector is obtained as the channel correlation between positions, normalised over the number of antennas.

The top-left square in Fig. \ref{fig:Antennas} shows the relative field strength for all the positions when MRT is applied to the centre location and only a single antenna is considered per sub-array. The beam towards the central user can not be distinguish as multiple locations (following a wave pattern) have a similar received field strength.

When the number of active antennas duplicates evenly per sub-array to a total of 16, the upper-left square shows an overall increase of the received field strength for all the locations, although the beam towards the reference point is yet not clear. However, this beam starts to be recognisable when 32 antennas are active. For 64 antennas, MRT achieves a clear spot on the target user. Even though \textit{RadioWeaves} is the topology which creates the richest multipath environment, the number of active antennas 
is vital to increase the beam directivity.

\section{Favourable Power leakage to Victim Users}
The principle behind the massive MIMO concept is the superposition of waves generated by a huge amount of antennas that can create tiny beams to serve users. The larger the number of antennas, the narrower the beam, and the more users that can be multiplexed in a small area. In real scenarios however, we are limited by the number of antennas, therefore, the beam is not as narrow as it can be in theory and part of the energy is leaked to other (\textit{idle}) users that we call victim users as they are exposed to unwanted radiation or experience interference.

\begin{figure}
    \centering
    \includegraphics[width=\linewidth]{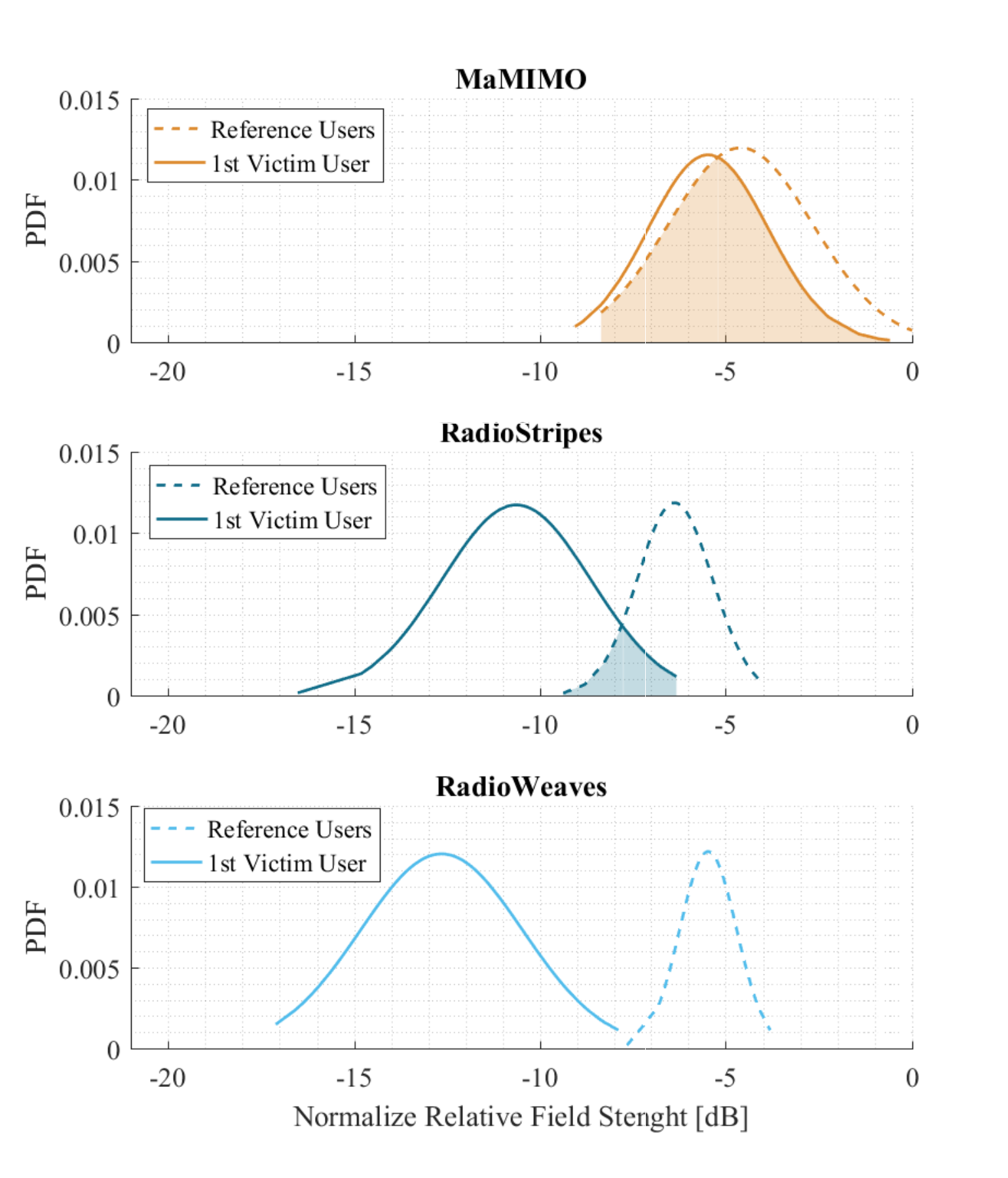}
    \caption{Weaving the antennas in the environment is expected to result in locally-confined communication, which means that it is possible to target a user precisely without harming other nearby users. We study the spatial confinement in our measured channel database by iteratively targeting each location in the dataset, and logging also the RSS measured in the most interfered location, which we call the victim user. Clearly, a \emph{RadioWeaves} weaving methodology results in the most spatial confinement, represented by the smallest overlap between the target and victim users signal strength histograms.}   
    \label{fig:PDF}
\end{figure}

Based on a dense channel database, with a distance difference between users of $150mm$, we quantify experimentally the array gain versus power leakage for different array topologies, as represented in Fig.~\ref{fig:PDF}. We apply MRT to a single user, identified as the "reference user", and measure the received power at that target user as well as all the victim users in the area. Fig.~\ref{fig:PDF} then plots the distribution of the received power in the reference user, obtained over all possible user locations in the dataset, and the distribution of the received power at the first victim user, which is the user experiencing the highest interference power in the measured area. The value of the relative field strength depends on the RSS value obtained in Section \ref{RSSI}, while the values related to the victim users relies on the beam directivity discussed in Section \ref{Sec. Antennas_MRT}.

From Fig.~\ref{fig:PDF} we see that the traditional \emph{Massive MIMO} array has the highest mean normalised power to the Reference User, which is expected as it also had the highest mean normalised RSS. However, we can see from the same figure that there is also a strong power leakage to the first strongest victim users (solid orange line). The impact of the RSS is also evident in the case of the \textit{RadioStripes} antenna distribution, which shows in the dashed blue line a lower mean field strength of the reference user, in comparison with the \textit{Massive MIMO} case, but also a reduction in the leakage of the victim user (solid blue line) due to the spatial diversity created by the antenna distribution. As a consequence of the antenna distribution and rich multi-path environment, the \emph{RadioWeaves} case presents in Fig. \ref{fig:PDF}, the lower leakage to the first victim user. 

In addition we calculate the probability of the victim's received field strength being higher than the reference. Below are those values for each configuration: 
\begin{itemize}

   \item \textit{Massive MIMO:} 18.3\%;
   \item \textit{RadioStripes:} 0\%;
   \item \textit{RadioWeaves:} 0\%.
\end{itemize}

In other words, the \textit{Radiostripes} and \textit{RadioWeaves}, in this experiment guarantee all target users a higher power than non-target ones. 

\section{High-accuracy User Positioning}
The large amount of information provided by the antennas in the proposed systems can be used to provide extra services to the users. One service in particular is of high interest: User localisation. Since the antennas of a \textit{RadioWeaves} system are used to focus wireless power in the spatial domain, the system has information about the position of the users.

In \cite{bast2019csibased}, the measured channels of the ultra-dense \textit{RadioWeaves} dataset were used to train a Machine Learning model to estimate the position of the users. The Machine Learning model is based on Convolutional Neural Networks (CNNs), which have been proven to be very effective to extract complex features of large amounts of data. In this case, the CNN used the CSI as input and was trained to estimate the exact location of the user. After training, the network was able to localise the users with an accuracy of around $20 mm$ and lower, depending on the scenario. This was achieved for both LoS and nLoS scenarios. 

\begin{figure}[!ht]
    \centering
    \includegraphics[width=\linewidth]{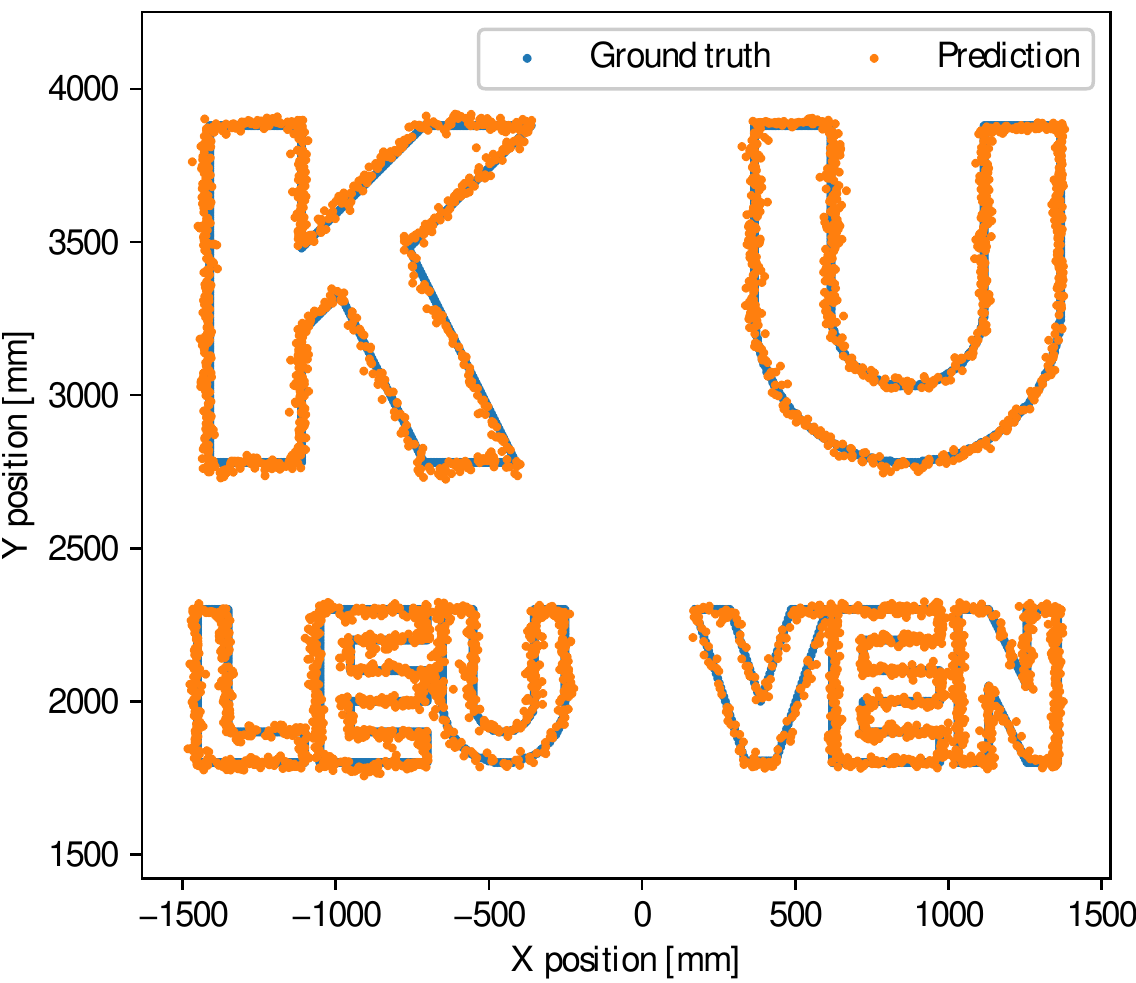}
    \caption{Illustration of the localisation system based on RadioWeaves. A Convolutional Neural Networked was trained on the ultra-dense dataset to localise the user based on its wireless channel. The model reached a mean error of 16.63 mm on this set of channels, spelling out the name of our university "KU Leuven".}
    \label{fig:kul_pos}
\end{figure}

To illustrate the high performance of this system, the letters of our university "KU Leuven" were spelled out in locations in the dataset. Next, the model was asked to estimate these locations based on the channel. It was able to do so with a mean error of 16.63 mm (0.145 wavelengths). The result can be seen in Fig. \ref{fig:kul_pos}.

\section{Future research directions}
This paper has studied favourable coverage and propagation conditions for multiple indoor antenna topologies by means of an ultra-dense channel database. In addition, the performance of the well-known MRT precoder has been analysed, and the power focusing performance was quantified by analysing the power received by reference and victim users. While the study touches multiple aspects of indoor distributed massive MIMO systems, there are still multiple aspects that should be studied more thoroughly with this dataset as the 
analyses of ZF or R-ZF precoders, user scheduling and power control.

Nevertheless, this experiment is a starting point to carry out many more experiments, which consider multiple indoor environments and deployment scenarios. Furthermore, novel datasets should be created where users are also deployed along the z-axis. An essential aspect of being studied is also the mobile and fixed users co-existence. For future experiments, absolute power and phase numbers should be considered. 

Finally, many more antenna topologies must be studied both theoretically and experimentally. Along with different metrics orientated to reliability and low-latency as time and frequency analysis, coherence time and bandwidth.

\section{Conclusion }
This paper presents a measurement based validation of multiple strategies for weaving antennas in an indoor environment. Concretely, we have compared Massive MIMO (centralised), RadioStripes (linear) and RadioWeaves (distributed) antenna topologies. It is clear that the more the antennas are distributed, the more favourable the coverage, propagation conditions and spot focusing performance become. Furthermore, all topologies achieve remarkable indoor localisation performance. A drawback of distributed topologies however, is that not all antennas contribute equally, and a reduced array gain towards a single user is obtained. 

In addition, this paper presents an ultra-dense dataset that is a strong alternative to idealistic channel models typically used in the community. This dataset allows to benchmark some typical statistical assumptions. Beyond that, it allows to emulate precoder performance for a realistic indoor scenario. This paper hence proposes a data-driven performance evaluation methodology, compared to traditional wireless approaches relying on statistical models. 

\section*{Acknowledgement}
We would like to thank Liesbet Van der Perre for her contributions to the RadioWeaves concept, and Fredrik Tufvesson and Ove Edfors for their help with the Massive MIMO testbed configuration. 
This work was funded by the European Union's Horizon 2020 under grant agreement no. 732174 (ORCA project) and by the Research Foundation Flanders (FWO) SB PhD fellowship, grant no. 1SA1619N.

\ifCLASSOPTIONcaptionsoff
  \newpage
\fi

\bibliographystyle{IEEEtran}
\bibliography{IEEEabrv,Distributed}

\end{document}